\documentclass[12pt]{svjour2}                    
\smartqed  
\usepackage{graphicx}
 \usepackage{mathptmx}      
\usepackage{latexsym}
\usepackage{amsfonts,amssymb,latexsym}
\usepackage{hyperref}
\begin{document}

\title{Observables for Brownian motion  on manifolds 
}


\author{Pavel Castro-Villarreal
}


\institute{P. Castro-Villarreal \at
             Centro de Estudios en F\'isica y Matem\'aticas B\'asicas y Aplicadas,  Universidad Aut\'onoma de Chiapas, C.P. 29050, Km. 8 Carretera Emiliano Zapata Tuxtla Guti\'errez, Chiapas, M\'exico\\[0.5em] 
              Tel.: +961 130 8913\\
              \email{pcastrov@unach.mx}           
           }

\date{Received: date / Accepted: date}

\maketitle

\begin{abstract} We study the geometrical influence on the Brownian motion over curved manifolds. We focus on the following intriguing question: what observables are appropriated to measure Brownian motion in curved manifolds? In particular, for those d-dimensional manifolds embedded in $\mathbb{R}^{d+1}$ we define three quantities for the displacement's notion, namely, the geodesic displacement, $s$, the Euclidean displacement, $\delta{\bf R}$, and the projected Euclidean displacement $\delta{\bf R}_{\perp}$. In addition, we exploit the Weingarten-Gauss equations in order to calculate the mean-square Euclidean displacement's in the short-time regime. Besides, it is possible to prove exact formulas for these expectation values, at all times, in spheres and minimal hypersurfaces. In the latter case, Brownian motion corresponds to the typical diffusion in flat geometries, albeit minimal hypersurfaces are not intrinsically flat. Finally, the two-dimensional case is emphasized since its relation to the lateral diffusion in biological membranes.\\

\noindent 05.40.Jc, 87.15.Vv, 02.40.Hw, 87.16.D
\end{abstract}

\section{Introduction}
\label{intro}

\noindent Brownian motion occurs as a representation of a plenty of phenomena arising in various contexts ranging from particle physics \cite{Quark},  general relativity \cite{smerlak} and condensed matter \cite{review}. In the last decade there has been much interest in the study of diffusive processes on manifolds, motivated by problems coming form biophysics \cite{Domanov}. The transport processes ocurring on a biological cell is an interesting and complex problem. In particular, the motion of an integral protein through the plasma membrane  has been approached from different point of views, which are themselves complementary. The most basic viewpoint is based on the Smoluchowski's equation of a punctual particle on  the membrane considered as two-dimensional regular curved surface. Further approaches consider membrane's  thermal fluctuations \cite{Gustaffson}-\cite{Seifert}, as well as dynamical fluctuations coupling to the stochastic motion of the protein \cite{Naji}, and finite-size effects of the protein \cite{Naji2}.

In this paper we use the Smoluchowski's approach to study the geometrical component on the diffusion processes of the integral protein on the membrane.  In general, we focus in the intriguing question: what observables are appropriate to measure Brownian motion in the present context?  By analogy with Euclidean spaces it is not difficult to realize that the actual motion of the particle is through geodesic displacements, thus, the  geodesic distance, $s$, is the proper notion for displacement on curved spaces \cite{Faraudo}, however, this quantity would be difficult to measure in an experiment. Nevertheless, unlikely Euclidean geometries for those curved manifolds embedded in $\mathbb{R}^{d+1}$ we have various quantities that also undergo stochastic dynamics, for instance,  the Euclidean displacement $\delta{\bf R}$, and the projected Euclidean displacement $\delta{\bf R}_{\perp}$.   In this way, albeit Euclidean displacements are not the physical displacements  they can be used as observables for the Brownian motion. Thus, from a practical point of view it would be easier to measure the Euclidean displacements than the geodesic displacement at least for Brownian motion on membranes (perhaps  for the Brownian motion on  curved space-time it is in the other way around because we are indeed immersed in that manifold). 

The Euclidean displacement has been, already, considered in  \cite{Muthukumar} for Guassian polymer wrapping curved interfaces and also in \cite{Holyst} to approach the transport modes on a cell membrane. The projected displacement has also been used  in \cite{Gustaffson} and \cite{Seifert}.  It is interesting that $\delta{\bf R}_{\perp}$, as well as the Euclidean displacement, is refering to the ambient space where the manifold is embedded. In this sense, the mean-square (MS) geodesic displacement, $s$, will have an influence from the intrinsic geometry whereas the Euclidean displacements, $\delta{\bf R}$ and $\delta{\bf R}_{\perp}$, will have an influence from the extrinsic geometry.  In particular, in this work we study the mean-square (MS) values for the Euclidean and projected displacements using the method developed at \cite{Castro-2010}. Also, this method allow us to give closed expressions for  these mean values for spherical and minimal hypersurfaces for all time values. It is shown that the diffusion on minimal hypersurfaces measured from the observable $\delta{\bf R}$ corresponds to the typical diffusion in agreement with the cubic minimal surfaces already studied in \cite{Holyst} and \cite{Anderson}.

This paper is organized as follows. In section 2, we summarized geometrical concepts used to describe intrinsic and extrinsic observables. In section 3, we present the diffusion equation and the displacement observables on curved manifolds and submanifolds. In section 4, we present the operator method used to evaluate the expectation values of observables. In particular, we study the short-time regime of the observables $\delta{\bf R}^2$, $\delta{\bf R}$ and $\delta{\bf R}^2_{\perp}$.  In addition, we show exact results for $\left<\delta{\bf R}^2\right>$ and $\left<\delta{\bf R}\right>$ for spheres and minimal hypersurfaces valid at all time values. Finally, in section 5, we summarized our main results and we give our concluding perspectives.  
\section{Preliminaries and notation}
\label{sec:1}

In this section we review the preliminary notions about  manifolds and sub-manifolds (following \cite{Nakahara} and \cite{Spivak}) needed to describe the observables for Brownian motion on curved manifolds.   Let $\mathbb{M}$ be a $d$-dimensional manifold and $U\subset\mathbb{M}$  a local neighborhood. By definition of manifold $U$ is locally diffeomorphic to a piece of Euclidean space.   In particular, we are interested in manifolds endowed with a Riemannian metric $g:T_{p}\left(\mathbb{M}\right)\times T_{p}\left(\mathbb{M}\right)\to\mathbb{R}$ given by ${\bf g}=g_{ab}~dx^{a}\otimes dx^{b}$, where  $g_{ab}$ is the meric tensor and $T_{p}\left(\mathbb{M}\right)$ is the tangent space for each $p\in\mathbb{M}$. Also, here, the Riemann tensor is denoted by ${\bf R}=R^{a~~}_{~bcd}~{e_{a}}\otimes dx^{b}\otimes dx^{c}\otimes dx^{d}$  and  the Ricci scalar curvature\footnote{The set $\left\{e_{a}\right\}$ is a basis for the tangent space and $dx^{a}$ is the corresponding basis in the dual tangent space.} by $R_{g}$. In addition, it is convenient to introduce the Laplace-Beltrami operator on scalars defined by $\Delta_{g}:C^{2}\left(\mathbb{M}\right)\to\mathbb{R}$ given by
\begin{eqnarray}
\Delta_{g}~\cdot=\frac{1}{\sqrt{g}}\partial_{a}\left(\sqrt{g}g^{ab}\partial_{b}~\cdot~\right),
\end{eqnarray}
where $g=\det{g_{ab}}$ and $g^{ab}$ is the inverse metric tensor. Also, the derivations are defined by $\partial_{a}=\partial/\partial{x}^{a}$, where $x^{a}$ with $a=1, \cdots, d$ are local coordinates of some patch in the manifold.

Let $\Sigma$ be a $d$-dimensional sub-manifold orientable in $\mathbb{R}^{d+1}$ with embedding functions ${\bf X}:\mathcal{D}\subset\mathbb{R}^{d}\to\Sigma\subset\mathbb{R}^{d+1}$ which assigns  $\left(u_{1},\cdots, u_{d}\right)\to{\bf X}\left(u_{1},\cdots,u_{d}\right)$.  Each vector of the tangent space $T_{p}\left(\Sigma\right)$  can be spanned by $\left\{{\bf e}_{a}\right\}$, where ${\bf e}_{a}:=\partial_{a}{\bf X}$ are the tangent vectors.  The 1$^{st}$ Fundamental Form of the submanifold  is defined by ${\rm\bf I}:T_{p}\left(\Sigma\right)\to \mathbb{R}$ given by ${\rm \bf I}\left(\bf v\right)={\bf v}\cdot {\bf v}$;  thus the metric tensor adopts the simple structure $g_{ab}={\bf e}_{a}\cdot{\bf e}_{b}$. Here,  $\cdot$ is the canonical inner product of $\mathbb{R}^{d+1}$ and $\left|~\cdot~\right|$ is the norm comming from this inner product.

The normal direction to the tangent space is determined by the Gauss map ${\bf N}:\Sigma\to S^{d}$ defined by ${\bf N}^{2}=1$ and ${\bf N}\cdot{\bf e}_{a}=0$ for each $a$. The curvature of the hypersurface can be understand in terms of the change of the Gauss map, thus the 2$^{nd}$ Fundamental Form is defined by ${\rm\bf II}:T_{p}\left(\Sigma\right)\to\mathbb{R}$ given by ${\rm \bf II}\left({\bf v}\right)=d{\bf N}_{p}\cdot{\bf v}$; here the components of this form are the extrinsic curvature tensor $K_{ab}={\bf e}_{a}\cdot\partial_{a}{\bf N}$.  The trace of this tensor is the mean curvature  $K=g^{ab}K_{ab}$. Also, it should be noted that the ``egregium" Gauss theorem implies that the Riemann tensor, $R_{abcd}\equiv K_{ac}K_{bd}-K_{ad}K_{bc}$, depends just on the intrinsic geometry.  The  tangent space changes direction for each point $p$ in the hypersurface. The manner how this change happens  is captured by the Weingarten-Gauss structure equations

\begin{eqnarray}
\label{WG1}
\nabla_{a}{\bf e}_{b}&=&-K_{ab}{\bf N},\\
\nabla_{b}{\bf N}&=&K_{b}^{~a}{\bf e}_{a},
\label{Weingarten-Gauss}
\end{eqnarray}
where $\nabla_{a}$ is the covariant derivative compatible with the metric $g_{ab}$.
\section{Diffusion equation and observables on curved manifolds}
\label{sec:2}
In this section, we introduce the simplest model to study Brownian motion on curved manifolds.  This is a direct generalization of the diffusion equation  on Euclidean spaces, which basically consist of replacing the Euclidean Laplacian by the Laplace-Beltrami operator $\Delta_{g}$. This operator is often used to describe how a substance diffuses over a curved manifold. Also we may think it as heat diffusing on manifolds or polymer confined to curved surfaces \cite{Muthukumar}. Furthermore, it can be studied to determine the quantum propagator of a free particle on curved spaces \cite{Chaichian}. 

For a single particle diffusion over a manifold, we are interested in the probability density  $P: \mathbb{M}\times\mathbb{M}\times\mathbb{R}^{+}\to\mathbb{R}^{+}$ such that $P\left(x,x^{\prime},t\right)dv$  means the probability to find a diffusing  particle in the volume element $dv$ when particle started to move at $x^{\prime}$ and it has passed a time t. This distribution is the same as the so called heat kernel as well as the propagator for a quantum particle moving on the manifold. In particular,  here, the diffusion is governed by the diffusion equation (for $P$)
\begin{eqnarray}
\frac{\partial P\left(x,x^{\prime},t\right)}{\partial t}=D\Delta_{g}P\left(x,x^{\prime},t\right),
\label{diff.Eq}
\end{eqnarray}
where $D$ is the diffusion coefficient. Also, we required that $P$ satisfies the initial condition at time $t\to0$
\begin{eqnarray}
 \lim_{t\to 0}P\left(x,x^{\prime},t\right)=\frac{1}{\sqrt{g}}\delta^{d}\left(x-x^{\prime}\right).
 \label{ini.cond}
\end{eqnarray}
In addition, the existence of $P$ (on a complete Riemannian manifold) is guaranteed if it also satisfies
\begin{eqnarray}
P\left(x,x^{\prime},t\right)&=&P\left(x^{\prime},x,t\right)
\label{simmetry}\\
P\left(x,x^{\prime},t\right)&=&\int_{\mathbb{M}}dv_{y} P\left(x,y,t-s\right)P\left(y,x^{\prime},s\right)
\label{convolution}
\end{eqnarray}
for any $s>0$. This is the content of the theorem by Schoen and Yau  \cite{Yau}. These properties are easier to understand using the quantum analogy. The symmetry property (\ref{simmetry}) means that the evolution of a quantum particle from $x$ to $x^{\prime}$ is the same evolution from $x^{\prime}$ to $x$, and the convolution property (\ref{convolution}) means that this propagation from $x$ to $x^{\prime}$ can be decomposed into individual propagations from $x$ to $y$ and $y$ to $x^{\prime}$, for each value of $y\in \mathbb{M}$. 

In mathematics one of the challenges consist of estimate expressions for the heat kernel or the probability density (see \cite{Grygorian} for a review) and one reason is because heat kernel encoded a strong connection with topological invariants of the manifold \cite{gilkey}. For most geometries, there is not a closed form of the probability distribution. However, for short times there is a formal series solution for $P\left(x,x^{\prime},t\right)$ in terms of the Minakshisundaram-Pleijel coefficients \cite{Pleijel}, which depends on both $x$ and $x^{\prime}$ \cite{Denjoe}; this series expansion is also called the parametrix expansion \cite{Lafferty}. It is noteworthy to mention that short-time depends on the especific geometrical dimensions of the manifold; below we will give a precise definition of what we mean by short-time in the present context. In what follows, we are interested in physical observables in order to have information about Brownian motion on manifolds.

In order to get some insight about  Brownian motion we often look at mean values of displacement and square displacement. These physical observables give us information of how particle diffuses in the space.  For the Brownian motion on curved manifolds we have, in addition,  other random variables that are useful to understand the phenomena.  These quantities  are continuos scalar functions $\mathcal{O}:\mathbb{M}\to\mathbb{R}$ defined on the manifold and, normally,  are  functions in $C^{\infty}\left({\mathbb{M}}\right)$, although  this is not the general case. The only rigid  condition for an observable $\mathcal{O}\left(x\right)$ is that its expectation value respect to $P$, defined in the standard fashion
\begin{eqnarray}
\left<\mathcal{O}\left(x\right)\right>=\int_{\mathbb{M}} dv~ \mathcal{O}\left(x\right)P\left(x,x^{\prime}, t\right),
\end{eqnarray}
is well-defined for all points $x^{\prime}$ in the manifold and for all time values. Note that $\left<\mathcal{O}\left(x\right)\right>$ depends  on the initial position ${ x}^{\prime}$. Since we are concerning about  physical observables for  Brownian motion on curved spaces, we would like to address the question about  what functions $\mathcal{O}\left(x\right)$ on the manifold are useful to describe Brownian motion. In particular, we refer to the displacement of a Brownian particle on the manifold. 

Let us recall that a free Brownian particle over an Euclidean space undergoes statistical fluctuations that are isotropic and homogenous. These properties imply rotational and translational symmetries of the Euclidean space. In addition, these symmetries appear already in the Laplacian of the diffusion equation. Furthermore, the probability density inherited this invariance. For this case,  the displacement is one observable of interest. The displacement is given by  $\delta{\bf R}={\bf X}-{\bf X}^{\prime}$, where  ${\bf X}$ and  ${\bf X}^{\prime}$ are  two vectors in the Euclidean space; ${\bf X}^{\prime}$ is the started point of the motion of the particle. Using the probability density for this case it is not difficult to show that   $\left<\delta{\bf R}\right>=0$,  and  $\left<\delta{\bf R}^{2}\right>=2dDt$; indeed, these observables capture the rotational and translational invariance. 
Now, the Brownian motion of a free particle over curved manifolds will inherit the symmetries of the manifold. The displacement in this case is given by the geodesic distance, defined as follows.
\paragraph{Geodesic displacement.} 
In the intrinsic point of view
,  one particle displaces from $p_{1}$ to $p_{2}$ in $\mathbb{M}$ throughout a differentiable curve $\gamma:I\subset\mathbb{R}\to\mathbb{M}$. Thus the  {\small\sc geodesic displacement} (GD) or geodesic distance, $\mathcal{O}_{1}\equiv s$, is defined to be the infimum length of geodesic curve beteween $\gamma\left(t_{1}\right)={\bf X}\circ\beta\left(t_{1}\right)$ and  $\gamma\left(t_{2}\right)={\bf X}\circ\beta\left(t_{1}\right)$, that is 
\begin{eqnarray}
s={\rm inf}\int^{t_{2}}_{t_{1}}\left|\gamma^{~\prime}\left({\bf X}\left(t\right)\right)\right|dt,
\label{gdistance}
\end{eqnarray}
where  ${\bf X}:\mathcal{D}\subset\mathbb{R}^{d}\to\mathbb{M}$ is a parametrization of the manifold and $\beta:I\subset\mathbb{R}\to\mathbb{R}^{d}$ is the pre-image of curve $\gamma$. The expectation value, $\left<s^2\right>$, captures geometrical data of the manifold, and it  gives the manner of how the intrinsic geometry influences Brownian motion. 

In addition, for those $d$-dimensional sub-manifolds, embedded in $\mathbb{R}^{d+1}$, we have Euclidean displacements defined as follows. 

\paragraph{Euclidean displacement.} In the extrinsic point of view
,  for Brownian motion over d-dimensional hypersurfaces ($\subset\mathbb{R}^{d+1}$), 
there is another observable that is referred to the ambient space $\mathbb{R}^{d+1}$. The {\small\sc Euclidean displacement} (ED)  is defined by $\delta{\bf R}:\mathbb{M}\to {\mathbb R}^{d+1}$ given by  $\delta{\bf R}={\bf X}-{\bf X}_{0}$ for  all points ${\bf X}\in \mathbb{M}$, where ${\bf X}$ is itself a parametrization of the hypersurface. This quantity also describes the displacement of the particle, but seeing it from the ambient space. Let us denote $\mathcal{O}_{2}\equiv\delta{\bf R}$ and $\mathcal{O}_{3}\equiv \left|\delta{\bf R}\right|^{2}$, where the distance function is  $\left|\delta{\bf R}\right|$.  

\paragraph{Projected displacement.}  Brownian motion also shows peculiar features in sub-spaces of the ambient space.  In particular, here, we are going to consider only projections of the hypersurface. The projected subspace  is defined by $\pi\left(\mathbb{R}^{d+1}\right)=\mathbb{R}^{d}$, where the projection map, $\pi:\mathbb{R}^{d+1}\to\mathbb{R}^{d}$, is defined as usual by $\pi\left({\bf V}\right)={\bf v}$ for ${\bf V}=\left({\bf v}, v_{0}\right)\in\mathbb{R}^{d+1}$.  In this subspace,  the  {\small{\sc projected displacement}} (PD) is defined by the composition map $\delta{\bf R}_{\perp}\equiv\pi\circ\delta{\bf R}$.  In particular, for a domain of $\mathbb{M}$ such that it can be covered with one coordinate neighborhood we are able to use the Monge parametrization ${\bf X}=\left({\bf x},h\left({\bf x}\right)\right)$, where  $h\left({\bf x}\right)$ is the height function for ${\bf x}\in \mathbb{R}^{d}$. In these terms we have   $\delta{\bf R}_{\perp}={\bf x}-{\bf x}_{0}$.  Let us denote $\mathcal{O}_{4}\equiv \left(\delta {\bf R}_{\perp}\right)^2$. Note that this observable is also an extrinsic measure of the displacement of the particle. 

\section{Expectation values of observables}
\label{sec:3}

The general problem is to find the mean values for  $\mathcal{O}_{1}$, $\mathcal{O}_{2}$, $\mathcal{O}_{3}$ and $\mathcal{O}_{4}$ for an arbitrary manifold and sub-manifold.  In principle, these expectation values can be eva\-lua\-ted through the formal series solution of the diffusion equation in terms of the Minakshisundaram-Pleijel coefficients mentioned above.  Here, we use an operator method  introduced at \cite{Castro-2010}. This method is inspired in the original calculations made by Perrin in his seminal papers about Brownian motion on spheres \cite{Perrin}. Next, $P$ is a density distribution satisfying (\ref{diff.Eq}), (\ref{ini.cond}), (\ref{simmetry}) and (\ref{convolution}). The method is encoded in the following result.

{\proposition\label{prop1}{\it Let $\mathcal{O}:\mathbb{M}\to \mathbb{R}$ be an observable either in $C^{\infty}\left(\mathbb{M}\right)$ or  ${\mathbb{R}^{d+1}}\times C^{\infty}\left(\mathbb{M}\right)$, then the expectation value of  $\mathcal{O}$, with respect to a probability density $P$,  have the following derivations respect to time
\begin{eqnarray}
\frac{\partial^{k}\left<\mathcal{O}\left(x\right)\right>}{\partial t^{k}}=D^{k}\int dv~ \Delta^{k}_{g}\mathcal{O}\left(x\right)P\left(x, x^{\prime}, t\right)+ D^{k}\int dv~\nabla_{a}J^{a}_{k},
\label{result0}
\end{eqnarray}
where
\begin{eqnarray}
J^{a}_{k}=\sum^{k}_{j=0}\left\{\left(\Delta^{k-j-1}_{g}\mathcal{O}\right)\nabla^{a}\Delta^{j}_{g}P-\left(\Delta^{k-j-1}_{g}P\right)\nabla^{a}\Delta^{j}_{g}\mathcal{O}\right\}.
\end{eqnarray}}}
The proof of this proposition is along the following lines. First,  let us differentiate $\left<\mathcal{O}\left(x\right)\right>$ with respect time, then substitute the diffusion equation (\ref{diff.Eq}). Next, we   use  the Green formula (\ref{GreenFormula}) and substitute the initial condition  (\ref{ini.cond}). We gave a proof in \cite{Castro-2010}, but this same result is also proved in \cite{Grygorian}. 

{\remark{ Note that $J^{a}_{k}$ for each $k$ is a vector field on $\mathbb{M}$ thus, by the divergence theo\-rem, for compact manifolds  the right hand side of Eq. (\ref{result0}) vanishes \cite{Chavel}, except for non-trivial topologies like circle $S^{1}$ or torus $T^{2}$; these cases will be analysed elsewhere in \cite{Castro-12.2}. For manifolds with boundaries we choose that  $P$ and $\nabla^{a}P$ vanish at the boundary therefore right hand side of Eq. (\ref{result0}) also vanishes. In particular, for these cases we have that 
\begin{eqnarray}
\left.\frac{\partial^{k}\left<\mathcal{O}\left(x\right)\right>}{\partial t^{k}}\right|_{t=0}=\left.D^{k}\Delta^{k}_{g}\mathcal{O}\left(x\right)\right|_{x=x^{\prime}}.
\end{eqnarray} }} In what follows, we will consider compact manifolds or manifolds where $P$ and $\nabla^{a}P$ vanish at the boundary. In addition, assuming that  $\left.\partial^{k}\left<\mathcal{O}(x)\right>/\partial t^{k}\right|_{t=0}$ are well defined on $\mathbb{R}^{+}$, for a given  $\mathbb{M}$, we define the remainder $R_{n}\left(t\right)$ by
\begin{eqnarray}
\left<\mathcal{O}\left(x\right)\right>=\sum^{n}_{k=0}\frac{G^{{\small \mathcal{O}}}_{k}}{k!}\left(Dt\right)^{k}+R_{n}\left(t\right),
\label{formula}
\end{eqnarray}
where the terms $G^{{\small \mathcal{O}}}_{k}\equiv\left.\Delta^{k}_{g}\mathcal{O}\right|_{x=x^{\prime}}$ are purely geometric factors. Thus by Taylor theorem (\ref{Taylor}) the remainder can be written in terms of the expectation value itself. The definition of the remainder, equation (\ref{formula}), is useful if we are able to prove that $\lim_{n\to \infty}R_{n}\left(t\right)=0$  because in this case $\left<\mathcal{O}\left(x\right)\right>$  has a series Taylor representation 
\begin{eqnarray}
\left<\mathcal{O}\left(x\right)\right>=\sum^{\infty}_{k=0}\frac{G^{{\small \mathcal{O}}}_{k}}{k!}\left(Dt\right)^{k}.
\label{formula1}
\end{eqnarray} 
It is notable that not all expectation values satisfies $\lim_{n\to \infty}R_{n}\left(t\right)=0$; in \cite{Castro-12.2} will be an example of this.  For a given observable  the difficulty lies to evaluate the terms $G^{{\small \mathcal{O}}}_{k}$ and to estimate $\lim_{n\to \infty}R_{n}\left(t\right)$.  The equation (\ref{formula}) is very useful to access the short-time regime of the Brownian motion for the general manifold case, but also it can be used to find closed formulas valid for all times for some specific manifolds.

\subsection{The Brownian motion at short-time regime }
 \vskip1em
\noindent 
In what follows, we are going to give some estimations for the mean values of the  observables$\mathcal{O}_{1}$, $\mathcal{O}_{2}$, $\mathcal{O}_{3}$ and $\mathcal{O}_{4}$ at the short-time regime, that is for times $t\sim\tau_{G}$, where $\tau_{G}=3d/\left|R_{g}\right|D$ is called geometrical time \cite{Castro-2012}.  Basically, we use the proposition (\ref{prop1}) and we  compute the factors $G^{O}_{k}$ for these observables with $k=1,2,3$.
\\
 \vskip1em
\subsubsection{Intrinsic observables on manifolds}

The mean-value of $s^2$ will capture intrinsic geometrical data of the manifold and it will give how this geo\-me\-try causes a change in the standard diffusive behaviour. The geometric factors $G^{\mathcal{O}_{1}}_{k}$ cannot be written, in general, in a closed form for each $k$. However, in \cite{Castro-2010} we have shown a formula for the mean-square geodesic displacement for the first three values $k=1,2,3$. Hence, this value can be written as
\begin{eqnarray}
\left<s^{2}\right>\approx 2dDt-\frac{2}{3}R_{g}\left(Dt\right)^{2}&+&\frac{1}{3!}\left[\frac{8}{15}R^{ab}R_{ab}\right.-\left.\frac{16}{45}R^{abcd}\left(R_{dbca}+R_{dcba}\right)\right.\nonumber\\
&&~~~~~~~~~~~~~~~~~-\left.\frac{16}{5}\left(\nabla^{a}\nabla^{b}+\frac{1}{2}g^{ab}\Delta_{g}\right)R_{ab}\right]\left(Dt\right)^{3}+\cdot\cdot\cdot\nonumber,\\
\label{mean-square}
\end{eqnarray}
where $\approx$  is defined through  the theorem (\ref{Teo1}). This result shows how the  mean-square GD is deviated from the planar expression by terms which are invariant under general coordinate transformations. As a consequence of the Gauss ``egregium'' theorem all these terms are isometric and  
are built with $O(d)$ invariant combinations of the Riemann tensor. 
In principle, this result is valid for every Riemannian manifold endowed with metric tensor $g_{ab}$.  Clearly, in a local neighborhood Brownian motion is not affected by the geometry of the manifold, but as far away as the particle reaches the boundary of this neighborhood the curvature effects become apparent.

{\example   On spheres $S^{d}$, where  Riemann curvature is $R_{abcd}=\frac{1}{R^{2}}\left(g_{ac}g_{bd}-g_{ad}g_{bc}\right)$, the expectation value of $s^{2}$ is given by    
\begin{eqnarray}
\left<s^{2}\right>\approx 2dDt-\frac{2}{3}\frac{d(d-1)}{R^{2}} \left(Dt\right)^{2}+\frac{4}{45}\frac{d(d-1)(d-3)}{R^{4}}\left(Dt\right)^{3}+\cdots.
\end{eqnarray}
 This result has also been obtained by direct calculation at \cite{Castro-2010}.}

{\example For regular surface embedded in $\mathbb{R}^{3}$, Riemann tensor components are $R_{abcd}=\frac{R_{g}}{2}\left(g_{ac}g_{bd}-g_{ad}g_{bc}\right)$, therefore the expectation value for these surfaces at short-time regime is given by
\begin{eqnarray}
\left<s^{2}\right>&\approx&4Dt-\frac{4}{3}K_{G}\left(Dt\right)^{2}-\frac{8}{15}\left[\frac{1}{3}K^{2}_{G}
+2\Delta_{g}K_{G}\right]\left(Dt\right)^{3}+\cdot\cdot\cdot,
\label{MSGD}
\end{eqnarray}
where $K_{G}\equiv 2R_{g}$ is the Gaussian curvature. Note that for developable surfaces, $K_{G}=0$, the mean-square geodesic displacement behaves like the typical diffusion in this short-time regime \cite{Faraudo}.}


\vskip3.5em
\subsubsection{Extrinsic observables on  submanifolds of $\mathbb{R}^{d+1}$}

\vskip0.5em
\noindent
For the Brownian motion over hypersurfaces of dimension $d$ (submanifolds of $\mathbb{R}^{d+1}$) 
we are interested to know the expectation value of  $\mathcal{O}_{2}\equiv\left|\delta{\bf R}\right|^{2}$.  In this case, also the geometric factors  for $\mathcal{O}_{2}$ cannot be evaluated in a closed form for a given integer $k$. Again, we are going to calculate the  factors $G^{\mathcal{O}_{2}}_{k}$ for $k=1,2,3$.  For instance,  for $k=1$ we have $G^{\mathcal{O}_{2}}_{1}=\Delta_{g}\left|\delta{\bf R}\right|^{2}=2\nabla_{a}\left(\delta{\bf R}\cdot{\bf e}^{a}\right)$. Recalling the trace of metric tensor, $g_{~a}^{a}=d$, and using  the Weingarten-Gauss equation, (\ref{Weingarten-Gauss}), we get
\begin{eqnarray}
G^{\mathcal{O}_{2}}_{1}=\left.\left(2d-2K\delta{\bf R}\cdot{\bf N}\right)\right|_{\delta{\bf R}=0}=2d
\end{eqnarray} 
In a similar way, by straightforward calculation, we obtain 
\begin{eqnarray}
G^{\mathcal{O}_{2}}_{2}&=&\left.\left[(\Delta_{g}K){\bf N}\cdot\delta{\bf R}-2K\nabla^{a}\left(K_{a}^{~b}{\bf e}_{b}\right)\cdot\delta{\bf R}-4\left(\nabla^{a}K\right)K_{a}^{~b}{\bf e}_{b}\cdot\delta{\bf R}-2K^{2}\right]\right|_{\delta{\bf R}=0}\nonumber\\&=&-2K^{2}\nonumber\\
\end{eqnarray}
and using geometrical identities in appendix (\ref{ap}) we get
\begin{eqnarray}
G^{\mathcal{O}_{2}}_{3}=2K^{2}K_{ab}K^{ab}-2K\Delta_{g}K-2\Delta_{g}\left(K^{2}\right)-4\nabla_{b}\left(K\nabla^{a}K_{a}^{~b}+\left(\nabla^{a}K\right)K_{a}^{~b}\right).\nonumber\\
\end{eqnarray}
Hence the mean-square Euclidean displacement $\left<\delta{\bf R}^{2}\right>$ is written as follows
\begin{eqnarray}
\left<\delta{\bf R}^{2}\right>\approx 2dDt-K^{2}\left(Dt\right)^{2}-\frac{1}{3}\left[K\Psi\left(K\right)+\Delta_{g}\left(K^{2}\right)+2\nabla_{b}J^{b}\left(K\right)\right]\left(Dt\right)^{3}+\cdots\nonumber\\
\label{resultado1}
\end{eqnarray}
where the scalar $\Psi\left(K\right)$ and vector $J^{a}\left(K\right)$ are defined as 
\begin{eqnarray}
\Psi\left(K\right)&=&\Delta_{g}K-KK_{ab}K^{ab},\nonumber\\
J^{a}\left(K\right)&=&K\nabla_{b}K^{ba}+2K^{ab}\nabla_{b}K.
\end{eqnarray}
This result shows how the mean-square ED is  deviated from the typical diffusion  behaviour. This deviation is also given by terms invariant under general coordinate transformations,  but now they are referred to the ambient space where hypersurface is embedded; they are also invariant under global rotations and traslations on the hypersurface. Like the geodesic displacement, in a local neighborhood, mean-square Euclidean displacement reproduces the standard Einstein kinematical relation. Also, when particle reaches the boundary of the neighborhood, the  curvature effects  emerge. For this case all curvature terms are built with $O(d)$-invariant of the se\-cond fundamental form or the extrinsic curvature tensor $K_{ab}$. In other words, this observable encoded extrinsic information of the hypersurface. 

Both observables, $s^{2}$ and $\left(\delta{\bf R}\right)^{2}$, reproduce the standard mean-square displacement at very short-times, $t\ll \tau_{G}$. It seems intuitive that at these short-times there are not at all curvature influences on the Brownian motion because the local neighborhood looks like Euclidean space. However, even for those times the xobservable $\mathcal{O}_{3}$  shows curvature effects on the Brownian motion.   Following the same procedure for $k=1,2,3$ we find
\begin{eqnarray}
\left<\delta{\bf R}\right>\approx-K{\bf N}Dt&-&\frac{1}{2}\left[\Psi{\bf N}+J^{b}{\bf e}_{b}\right]\left(Dt\right)^{2}\nonumber\\
&-&\frac{1}{3!}\left\{\left[\Delta_{g}\Psi-\Psi K_{cd}K^{cd}\right.\right.
-\left.\left.\left(2K_{cd}\nabla^{c}J^{d}+J^{b}\nabla_{a}K^{a}_{~b}\right)\right]{\bf N}\right.\nonumber\\&+&\left.\left[-2\nabla^{c}\Psi K_{c}^{~d}-\Psi\nabla_{a}K^{ad}\right.\right.
+\left.\left.J^{b}K_{cb}K^{cd}-\Delta_{g}J^{b}\right]{\bf e}_{b}\right\}\left(Dt\right)^{3}\nonumber\\
\end{eqnarray}
\noindent Therefore for non-zero times $t\ll \tau_{G}\equiv 3/2\left|R_{g}\right|D$  there is still a  contribution from the curvature on $\left<\delta{\bf R}\right>$. At short-times the normal direction of $\left<\delta{\bf R}\right>$ is  explained as follows. Since there is not any preferential direction, tangent components of $\delta{\bf R}$ cancel out in average and then, by symmetry, $\left<\delta{\bf R}\right>$ is along the normal direction. Nevertheless, as soon as the particle reaches the boundary of the local neighboorhood tangent components also contribute.

{\remark\label{rm2} For minimal hypersurfaces embedded in $\mathbb{R}^{d+1}$ the mean curvature $K$ is zero thus  $\Psi\left(K\right)=0$ and $J^{a}\left(K\right)=0$ then the expectation values $\left<\delta{\bf R}\right>$ and $\left<\delta{\bf R}^2\right>$ appear to be the same as those for the Brownian motion on flat spaces whereas expectation value of $s^2$  shows an influence of the curvature, since $R_{g}<0$.  }

{\example\label{ex1}l For a $d$-dimensional sphere $S^{d}$ of radius $R$, the components of the second fundamental form are $K_{ab}=\frac{1}{R}g_{ab}$, thus the mean curvature is $K=d/R$.  Then, we have $\Psi(K)=-d^2/R^3$ and $J^{a}(K)=0$. The started point in this case is ${\bf X}_{0}=R{\bf N}$. Therefore the expectation values of $\delta{\bf R}$ and $\delta{\bf R}^{2}$
are
\begin{eqnarray}
\label{dR}
\left<\delta{\bf R}\right>&\approx&-\frac{dDt}{R}\left(1-\frac{\left(dDt\right)}{2R^2}+\frac{\left(dDt\right)^{2}}{6R^4}+\cdots\right){\bf N}\\
\left<\delta{\bf R}^{2}\right>&\approx&2dDt-\frac{d^{2}}{R^{2}}\left(Dt\right)^{2}+\frac{d^3}{3R^4}\left(Dt\right)^{3}+\cdots.
\label{dR2}
\end{eqnarray}
This shows a different behavior in comparison to the one found using observable $s$.}

\subsubsection{Extrinsic observable in the projected subspace}

In the subspace $\mathbb{R}^{d}$,  the projected displacement was defined by $\delta{\bf R}_{\perp}\equiv\pi\left({\bf X}\right)={\bf x}$, for ${\bf x}_{0}=0$.  In order to compute the mean-square PD,  $\left<\delta{\bf R}^{2}_{\perp}\right>$, let us write $\mathcal{O}_{4}\equiv {\delta\bf R}^{2}_{\perp}=\mathcal{O}_{2}-h^{2}$ and $\delta{\bf R}_{\perp}={\bf X}-h\hat{\bf k}$, therefore the geometric factors satisfy 
\begin{eqnarray}
G^{\mathcal{O}_{4}}_{k}=G^{\mathcal{O}_{2}}_{k}-\left.\Delta^{k}_{g}h^{2}\right|_{{\bf X}=0}.
\end{eqnarray}
Thus, for the geometric factors with $k=1,2,3$ we have to calculate $\Delta_{g}h^{2}$, $\Delta^{2}_{g}h^{2}$ and $\Delta^{3}_{g}h^{2}$. For $k=1$, we have $\Delta_{g}h^{2}=2\nabla_{a}h\nabla^{a}h$. Using the expression for the metric in this parametrization, latter factor can be written as $\Delta_{g}h^{2}=2\frac{\left(\partial h\right)^{2}}{1+\left(\partial h\right)^{2}}$ and using the normal vector ${\bf N}$ in this parametrization we find
\begin{eqnarray}
\Delta_{g}h^{2}=2\left(1-N^{2}_{z}\right).
\end{eqnarray}
The terms $\Delta^{2}_{g}h^{2}$ and $\Delta^{3}_{g}h^{3}$ are left expressed in covariant form. By straighforward calculation we find
\begin{eqnarray}
\Delta^{2}_{g}h^{2}=4\left(\nabla^{a}\nabla^{b}h\right)\left(\nabla_{a}\nabla_{b}h\right)+4\left(\Delta_{g}h\right)^{2}+12\left(\Delta_{g}\nabla_{a}h\right)\left(\nabla^{a}h\right),
\end{eqnarray}
and
\begin{eqnarray}
\Delta^{3}_{g}h^{2}&=&8\left(\nabla^{a}\nabla^{b}\nabla^{c}h\right)\left(\nabla_{a}\nabla_{b}\nabla_{c}h\right)+32\left(\Delta_{g}\nabla^{a}\nabla^{b}h\right)\left(\nabla_{a}\nabla_{b}h\right)+32\Delta^{2}_{g}\left(\nabla_{a}h\right)\left(\nabla^{a}h\right)\nonumber\\
&+&10\Delta_{g}\left(\Delta_{g}h\right)^{2}-8\left(\Delta^{2}_{g}h\right)\left(\Delta_{g}h\right).
\end{eqnarray}
Note that $\nabla_{a}$ is the covariant derivative compatible with the metric $g_{ab}$ and therefore it itself depends on the height function $h$. As in the previous cases, the mean-square projected displacement is written as 
\begin{eqnarray}
\left<\delta{\bf R}^{2}_{\perp}\right>\approx 2\left(d-1+N^{2}_{z}\right)Dt+\frac{1}{2!}G^{\mathcal{O}_{4}}_{2}\left(Dt\right)^{2}+\frac{1}{3!}G^{\mathcal{O}_{4}}_{3}\left(Dt\right)^{3}+\cdots
\label{projected}
\end{eqnarray}
For very short-times $t\ll \tau_{G}$, mean-square projected displacement has the typical diffusion behavior in flat geometries $\left<\delta{\bf R}^{2}_{\perp}\right>=2dD_{proj}t$, but with a new diffusion coefficient 
\begin{eqnarray}
D_{proj}=\frac{D}{d}\left(d-1+N^{2}_{z}\right),
\label{Diff}
\end{eqnarray}
modified by a determined geometrical content. This means that for the Brownian motion observed from the projected sub-space the diffusion is reduced, since $D_{proj}$ is smaller than $D$. This modification is just a geometrical effect due to the point of view from where  Brownian motion is seen. However,  we can always rotate the hypersurface such that $N_{z}=1$ at ${\bf x}={\bf x}_{0}$; after such a rotation we get $D_{proj}=D$. 

{\remark The result (\ref{Diff}) has been obtained by different methods for two-di\-men\-sio\-nal surface at  \cite{Gustaffson} and \cite{Seifert} within the context of lateral diffusion of integral proteins in biomembranes.  In these works they also consider  the ther\-mal fluctuations of the membranes. For instance at \cite{Seifert}, under the basis of the Helfrich-Canham model \cite{Helfrich-canham} for fluid membranes it is computed the effective value of the diffusion coefficient when thermal fluctuations are considered. It would be interesting to evaluate the next contributions of order $t^{2}$ \cite{Castro-12.2}. }

\subsection{The Brownian motion for all time values}
\vskip0.5em
\noindent

We consider now the whole  series (\ref{formula}) for a particular set of observables. For these ob\-ser\-va\-bles, as we shall see,   we are able to give exact and closed formulae valid for all time values. Let us start we the following

{\proposition\label{prop2} Let  $\mathcal{O}\left(x\right)$ be an eigenfunction of Laplace-Beltrami operator  $\Delta_{g}$ with eigenvalue $-\lambda$, then the expectation value of $\mathcal{O}\left(x\right)$ is given by
\begin{eqnarray}
\left<\mathcal{O}\left(x\right)\right>=\mathcal{O}\left(x^{\prime}\right)\exp\left(-\lambda Dt\right)
\label{res1}
\end{eqnarray}}
 {\proof It is clear that  $\mathcal{O}\left(x\right)$ fulfill all conditions: $\mathcal{O}\left(x\right)$ is a differentiable function. Indeed, the $k$-th action of $\Delta_{g}$ is given by  $\Delta^{k}_{g}\mathcal{O}\left(x\right)=\left(-\lambda\right)^{k}\mathcal{O}\left(x\right)$. The remainder $R_{n}\left(t\right)$ have the following expression
 \begin{eqnarray}
 R_{n}\left(t\right)=\frac{\left(-\lambda D\right)^{n+1}}{n!}\int^{t}_{0}d\tau \left<\mathcal{O}\left(x\right)\right>\left(t-\tau\right)^{n}d\tau, 
 \end{eqnarray}
 then we have
 \begin{eqnarray}
\left| R_{n}\left(t\right)\right|&\leq& \frac{\left(-\lambda D\right)^{n+1}}{n!}\left|\int^{t}_{0}d\tau \left<\mathcal{O}\left(x\right)\right>\right|\left|\int^{t}_{0}\left(t-\tau\right)^{n}d\tau\right|\nonumber\\
&=& M\frac{\left(-\lambda D\right)^{n+1}}{\left(n+1\right)!}, 
 \end{eqnarray}
where $M$ is a number independent of $n$. It is elemantary that  for  any number $a$ and $\epsilon>0$ we have $a^{n}/n!<\epsilon$  for  sufficently large value of $n$, therefore $\lim_{n\to\infty}R_{n}\left(t\right)=0$. Now, using (\ref{formula}) we  get the wished result (\ref{res1}). \qed}

{\proposition\label{prop3} Let  $\mathcal{O}\left(x\right)\in C^{\left(2\right)}\left(\mathbb{M}\right)$ such that $\Delta_{g}\mathcal{O}\left(x\right)=C$ for each point on $\mathbb{M}$, where $C$ is a non-zero real constant, then the expectation value of $\mathcal{O}\left(x\right)$ is given by
\begin{eqnarray}
\left<\mathcal{O}\left(x\right)\right>=\mathcal{O}\left(x^{\prime}\right)+CDt
\label{res2}
\end{eqnarray}}
 {\proof It is clear that  $\mathcal{O}\left(x\right)$ fulfill all conditions of proposition (\ref{prop1}): $\mathcal{O}\left(x\right)$ is a differentiable function. Indeed, the $k$-th action of $\Delta_{g}$ is given by  $\Delta^{k}_{g}\mathcal{O}\left(x\right)=0$ for $k>1$. In this case, the remainder satisfies $\lim_{n\to \infty}R_{n}\left(t\right)=0$.  Now, using (\ref{formula}) we  get the wished result (\ref{res2}). \qed}

Motivated by last two propositions let us open the following questions.
Let ${\bf X}$ be a pa\-ra\-me\-tri\-za\-tion for an Euclidean sub-manifold and let $s$ be the geodesic distance in a manifold. Last result leads us to pose the following questions for $\lambda$ real,  {\bf 1.} What is the collection of $d$-dimensional sub-manifolds of $\mathbb{R}^{d+1}$ such that each submanifold has at least one parametrization ${\bf X}$ satisfying each of the following conditions ({\it i}) $\Delta_{g}{\bf X}=\lambda {\bf X}$ or ({\it ii}) $\Delta_{g}{\bf X}^{2}=\lambda {\bf X}^{2}$ or
({\it iii}) $\Delta_{g}{\bf X}^{2}=\lambda $? {\bf 2.} What is the collection of $d$-dimensional manifolds such that each of them satisfies the following conditions ({\it j}) $\Delta_{g}s^{2}=\lambda $ or ({\it jj}) $\Delta_{g}s^{2}=\lambda s^{2}$ in at least one local neighborhood? Answer of first question with condition ({\it i}) is given by the following two propositions.
{\proposition\label{prop4} {\it  The submanifold $\mathbb{M}$, embedded in $\mathbb{R}^{d+1}$, is a $d$-dimensional sphere $S^{d}$, with radius $R$,  if and only if there is  a pa\-ra\-me\-tri\-za\-tion ${\bf X}$ such that $-\Delta_{g}{\bf X}=\frac{d}{R^2}{\bf X}$, for non-zero $R$.}}

{\proof On one hand, let us assume that the submanifold $\mathbb{M}$ is part of a  $d$-dimensional sphere $S^{d}$, with radius $R$, then there is a parametrization ${\bf X}$ such that ${\bf X}=R{\bf N}$, where ${\bf N}$ is the normal vector point outward the hypersphere. Also, the mean curvature of $S^{d}$ is $K=d/R$. By the Weingarten-Gauss equations $\Delta_{g}{\bf X}=-K{\bf N}$, therefore  $-\Delta_{g}{\bf X}=\frac{d}{R^2}{\bf X}$. On the other hand, if $-\Delta_{g}{\bf X}=\lambda{\bf X}$ by the Weingarten-Gauss equations we get ${\bf X}=\frac{K}{\lambda}{\bf N}$ that is ${\bf X}^{2}=\frac{K^2}{\lambda}$, therefore  $\nabla_{a}\left(K^{2}/\lambda\right)={\bf X}\cdot{\bf e}_{a}=0$ $\Leftrightarrow$ $K$ is constant $\Leftrightarrow$ $\mathbb{M}$ is $S^{d}$.\qed}

{\proposition\label{prop5} {\it  The submanifold $\mathbb{M}$, embedded in $\mathbb{R}^{d+1}$, is a $d$-dimensional minimal hypersurface, $K=0$, if and only if there is  a parametrization ${\bf X}$ such that $-\Delta_{g}{\bf X}=0$. }}

{\proof On one hand, let us assume that the sub-manifold $\mathbb{M}$ is a $d$-dimensional minimal hypersurface, namely mean curvature vanishes, $K=0$, then by Weingarten-Gauss equations there is one parametrization such that  $\Delta_{g}{\bf X}=0$. On the other hand if there is a parametrization such that $\Delta_{g}{\bf X}=0$, then by the Weingarten-Gauss equations $K{\bf N}=0$, that is $K=0$.\qed}

We ignore the answer of first question for the conditions ({\it ii}) and ({\it iii}). However,  for condition ({\it ii}) it is clear that  spheres are examples for $\lambda=0$. Also, it is noteworthy to mention that condition ({\it ii}) is equivalent to the relation $2d-2K{\bf X}\cdot{\bf N}=\lambda{\bf X}^{2}$, thus minimal hypersurfacess do not belong to the collection of this condition. Also, minimal hypersurfaces belong to the collection associated to the condition ({\it iii}) since for these sub-manifols we have $\Delta_{g}{\bf X}^{2}=2d$. In addition, we ignore the answer of question two, but for condition ({\it j})  it is clear that at least flat geometries  are examples and for condition ({\it jj}) we do not know even if there is a manifold that satisfying $\Delta_{g}s^{2}=\lambda s^{2}$.  All, these results are useful to proof a general structure for expectation values of $\delta{\bf R}$ and $\delta{\bf R}^{2}$ for  minimal hypersurfaces  ($K=0$) and spheres $S^{d}$.  These results are 
encoded in the following theorems.

{\theorem\label{teo2} The expectation values of $\delta{\bf R}$ and $\delta{\bf R}^{2}$, with respect to $P$, for each minimal hyper surfaces   of dimension $d$ are given by
\begin{eqnarray}
\left<\delta{\bf R}\right>&=&0,\\
 \left<\delta{\bf R}^{2}\right>&=&2dDt, 
\end{eqnarray} 
for all values of time $t$.} 
{\proof Since the sub-manifold is a minimal hypersurface then by proposition (\ref{prop5}) we have $\Delta_{g}{\bf X}=0$ and by proposition (\ref{prop2}) we get $\left<{\bf X}\right>={\bf X}_{0}$ $\Leftrightarrow$ $\left<\delta{\bf R}\right>=0$. For the mean-square ambient displacement, let ${\bf X}$ a parametrization of the minimal hypersurface then by the Weingarten-Gauss equations $\Delta_{g}{\bf X}^{2}=2d$, therefore for the proposition (\ref{prop3}) we get $\left<{\bf X}^{2}\right>={\bf X}^{2}_{0}+2dDt$, then  $\left<\delta{\bf R}^{2}\right>=2dDt$. \qed}

{\remark This theorem (\ref{teo2}) is consistent with the general formula at the short-time regime  (\ref{resultado1}); see remark (\ref{rm2}). This is in agreement with the result for the cubic minimal surfaces explicitly studied in \cite{Holyst} and \cite{Anderson}.}

{\theorem\label{teo1} The expectation values of $\delta{\bf R}$ and $\delta{\bf R}^{2}$, with respect to $P$, for spheres $S^{d}$ are given by
\begin{eqnarray}
\label{dRt}
\left<\delta{\bf R}\right>&=&{\bf X}_{0}\left(\exp\left(- \frac{d}{R^{2}}Dt\right)-1\right),\\
 \left<\delta{\bf R}^{2}\right>&=&2R^{2}\left(1-\exp\left(-\frac{d}{R^{2}}Dt\right)\right),
\label{dR2t}
\end{eqnarray} 
for all values of time.} 
{\proof Since the sub-manifold is a sphere $S^{d}$ by proposition (\ref{prop4}) there is a pa\-ra\-me\-tri\-za\-tion ${\bf X}$ such that $-\Delta_{g}{\bf X}=\frac{d}{R^2}{\bf X}$, that is ${\bf X}$ is an eigenfunction of $\Delta_{g}$ with eigenvalue $-\frac{d}{R^2}$, therefore by proposition (\ref{prop2}) we get $\left<{\bf X}\right>={\bf X}_{0}\exp\left(-\frac{d}{R^2}Dt\right)$, where ${\bf X}_{0}$ is the started point. Now, since the sub-manifold is an sphere, thus $\Delta_{g}{\bf X}^{2}=0$, then $\left<\delta{\bf R}^{2}\right>=\left<{\bf X}^{2}\right>+\left<{\bf X}^{2}_{0}\right>-2\left<{\bf X}\right>\cdot{\bf X}_{0}$, therefore by proposition (\ref{prop2}) and (\ref{prop3}) we get the wished result. \qed}

Note that equations (\ref{dRt}) and (\ref{dR2t}) at the short-time regime reproduce (\ref{dR}) and (\ref{dR2}), respectively; see example (\ref{ex1}). Last result has been already found at \cite{Muthukumar} and \cite{Holyst} by alternative methods. Now, let us consider a d-dimensional infinite cylinder with radius $R$. This cylinder can be thought as ${\rm Cyl}\equiv S^{d-1}\times{\mathbb{R}}$. In this case, the embedding functions can be written in terms of that $(d-1)$-dimensional sphere  ${\bf X}=\left({\bf X}_{S^{d-1}}, z\right)$, where  $z\in\mathbb{R}$. The metric  can be written as $g_{ab}={\rm diag}\left(1, g_{ij}\right)$, where $g_{ij}$ being the metric of $S^{d-1}$ and the Laplace-Beltrami operator in this case is given by $\Delta_{{\rm Cyl}}=\Delta_{S^{d-1}}+\partial^{2}/\partial z^{2}$. The started point is chosen to be ${\bf X}_{0}=\left(1, 0, \cdots,0\right)$, thus we have the following 

{\corollary The expectation values of $\delta{\bf R}$ and $\delta{\bf R}^{2}$, with respect to $P$, for cylinders ${\rm Cyl}$ are given by
\begin{eqnarray}
\left<\delta{\bf R}\right>&=&{\bf X}_{0}\left(\exp\left(- \frac{d}{R^{2}}Dt\right)-1\right),\\
\left<\delta{\bf R}^{2}\right>&=&2Dt+2R^{2}\left(1-\exp\left(-\left(n-1\right)\frac{Dt}{R^{2}}\right)\right)
\end{eqnarray}  
for all values of time.}

{\remark In particular, we have $\left<\delta z\right>=0$ and $\left<\delta z^{2}\right>=2Dt$ which corresponds to the one dimensional diffusion.

{\example Let us take a hemisphere of radius $R$ and let  $\Pi\cong \mathbb{R}^{2}$ be the projected subspace from this hemisphere. Thus a parametrization of this hemisphere is  ${\bf X}:\Pi\to\mathbb{R}^{3}$ defined by 
\begin{eqnarray}
{\bf X}\left(\varphi, \rho\right)=\left(\rho\cos\varphi, \rho\sin\varphi, \sqrt{R^2-\rho^{2}}\right).\end{eqnarray} This means that metric tensor comes in the following form  $g_{ab}=diag\left(\rho^2, R^2/\left(R^2-\rho^2\right)\right)$ and the projected displacement becomes $\delta{\bf R}_{\perp}=\left(\rho\cos\varphi, \rho\sin\varphi, 0\right)$. In particular, we can verify that $\Delta_{g}\left(\delta{\bf R}^{2}_{\perp}\right)=4-\frac{6}{R^2}\rho^2$ for all values of $\rho\in\mathbb{R}^{+}$, thus it is not difficult to show that by a straighforward calculation we find
$\left.\Delta^{k}_{g}\left(\delta{\bf R}^{2}_{\perp}\right)\right|_{\rho=0}=4\left(-\frac{6}{R^2}\right)^{k-1}$ for all naturals $k\neq 0$. Also, for this observable, $\delta{\bf R}^{2}_{\perp}$, we can verify that the remainder $R_{n}\left(t\right)$ goes to zero for large $n$. Therefore in this regime the mean-square projected displacement is given by
\begin{eqnarray}
\left<\delta{\bf R}^2_{\perp}\right>=\frac{2}{3}R^2\left(1-e^{-\frac{6Dt}{R^2}}\right),
\label{dR2perp}
\end{eqnarray}
 \begin{figure}[h]
 \begin{center}
\includegraphics[width=0.8\linewidth]{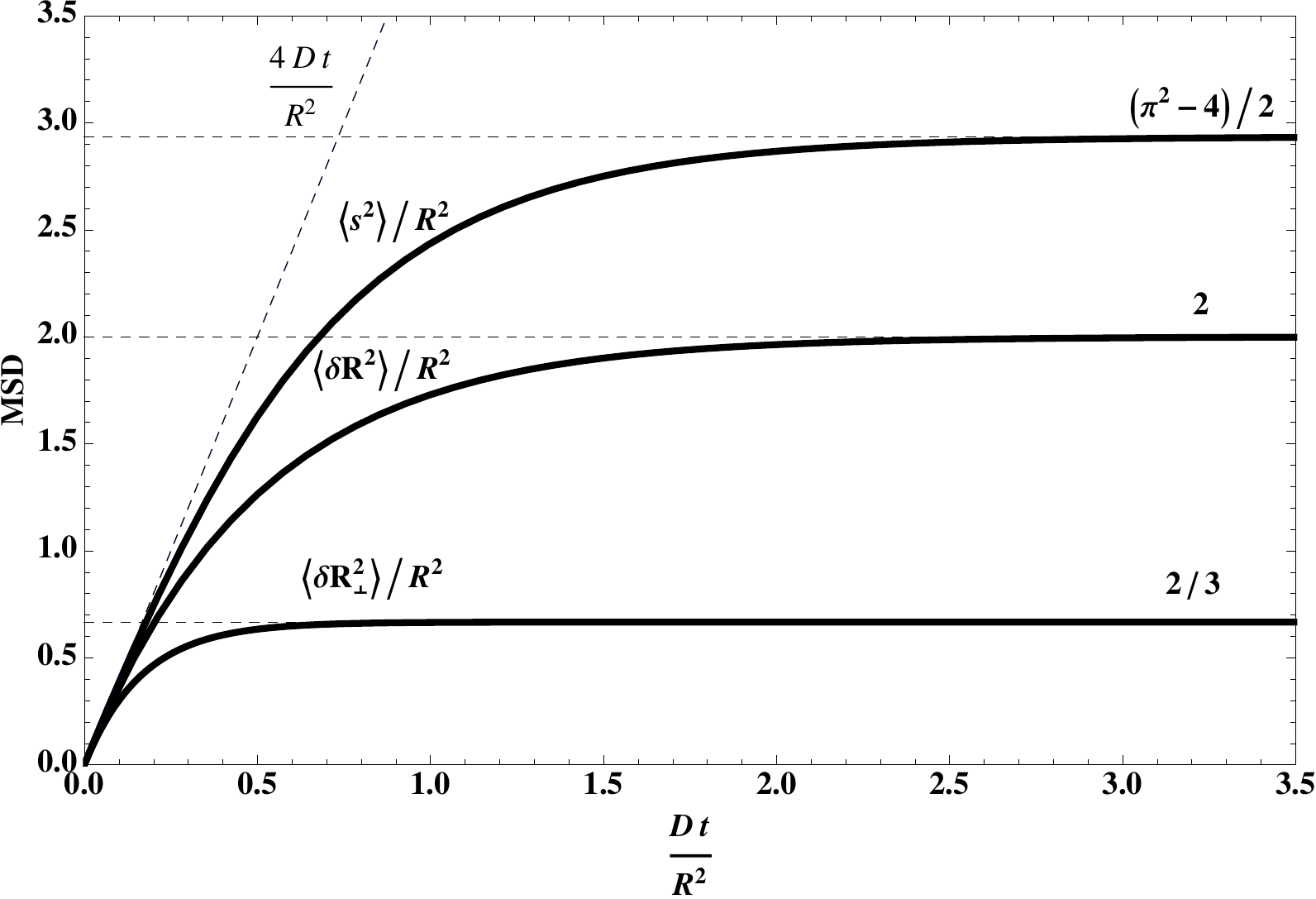}
 \label{fig1}
 \caption{{\small Mean-square geodesic and Euclidean displacements as a function of time for a Brownian particle diffusing on the sphere. The expectation value $\left<s^2\right>$ corresponds to our result in \cite{Castro-12.2}; the expectation value $\left<\delta{\bf R}^{2}\right>$ corresponds to the equation (\ref{dR2t}) and the expectation value $\left<\delta{\bf R}^{2}_{\perp}\right>$ corresponds to Eq. (\ref{dR2perp}). The straight lines stand for the short and long-time limits. }}
  \end{center}
 \end{figure} 
valid at all time values. For short-times this expression reproduces the general structure for the projected mean-square displacement (\ref{projected}). In particular, the diffusion coefficient does not change since $N_{z}=1$ at $\rho=0$. 

In figure (\ref{fig1}), we show a comparison between  expectation values of $\left<s^2\right>$, $\left<\delta{\bf R}^{2}\right>$ and $\left<\delta{\bf R}^2_{\perp}\right>$ for all time values;  these three observables coincide to  $4Dt$ for short-times, however, for large-times it is shown that the mean-square PD saturates faster than mean-square GD and ED do  as a consequence of the projection to the plane. It is also interesting that the saturation value (for large-times) goes to the value $\left<\delta{\bf R}^2_{\perp}\right>=\left(2/3\right) R^2$ rather than $2R^2$ as it happens for the observable $\left<\delta{\bf R}^{2}\right>$.   }

\section{Concluding perspective }

In this paper we have studied Brownian motion over curved manifolds and, mainly,  on Euclidean sub-manifolds of dimension d.
Our framework is based on the diffusion equation on curved manifolds. Here we have addressed the question about what functions $\mathcal{O}\left(x\right)$ are useful to described Brownian motion on curved spaces. In particular, we reviewed the notion about displacement and we define three  quantities for this notion, namely,  the geodesic displacement $s$, the Euclidean displacement, $\delta{\bf R}$, and the projected displacement, $\delta{\bf R}_{\perp}$. It is noteworthy to mention that the controversy posed in \cite{Faraudo} about the displacement is figure out by consider that all these displacement capture  information of the Brownian motion. Furthermore, from practical point of view it would be easier to measure extrinsic displacements than the geodesic one  because of the embedding of the membrane \cite{Seifert}.  

Here, we study the short-time regime of expectation values of $\delta{\bf R} $, $ \delta{\bf R}^2 $, and $ \delta{\bf R}^2_{\perp} $ using the operator method introduced at \cite{Castro-2010}. It is remarkable that for sufficiently small times, $ t \ll \tau_ {G} $, the quantities  $ \left <\delta {\bf R}^2\right> $ and $ \left< s^2\right> $ show no curvature effects while $\delta {\bf R}$ shows them.  Indeed, at this short-time regime the mean value for $\delta {\bf R}$ is proportional to the scalar curvature $K$ and it is pointing outward the hypersurface at the normal direction ${\bf N}$. In other words, it is like if local curvature has been manifesting  through a normal force acting on the particle. For the projected distance $\delta {\bf R}_{\perp}$ it is found that  diffusive coefficient $ D $ is modified by the projected one, $ D_ {\rm proj}\equiv \frac{D}{d}\left(d-1+N^{2}_{z}\right)$, by purely geometrical effects. However, we point it out that for a fix geometry we can always find a global rotation
such that $ D = D_ {\rm proj} $. Nevertheless, this is not the case for membranes under thermal fluctuations because, in general,  average of $N^{2}_{z}$ is not going to be equal to one \cite{Seifert}. In addition, it is shown that expectation values of eigenfunctions of Laplace-Beltrami operator can be written in a closed form. Particularly, it  is  proved that expectation values are linear in time for functions $\mathcal{O}$ that satisfy $\Delta_g \mathcal{O}={\rm constant}$. These results allow us to find closed expressions for expectation values corresponding to spheres and  minimal hypersurfaces. In particular, we found surprising that for the latter case the mean-square Euclidean displacement is exactly the one found for flat geometries, although  the Brownian motion has intrinsic curvature effects. A very similar effect happen for developable surfaces (with zero Gaussian curvature) at least for short-times \cite{Faraudo}.

For the future work, on one hand the surprise found for the minimal hypersurfaces using $ \delta{\bf R} $ immediately generalizes to higher order momenta. 
Even, it opens the possibility of finding an exact result for the density probability where Weierstrass-Enneper representation may play a central role. On the other hand,  the behaviour of the mean-square displacement (either intrinsic or extrinsic) for fluid membranes (i.e. described, for instance,  by the Helfrich-Canham model) under thermal fluctuations can be study, relatively simple, at least to the one-loop order in Feynman diagrams. 

\begin{acknowledgements}
\vspace{.5cm} We thank Metteo Smerlak,  Sendic Estrada Jim\'enez, and   Ramon Casta\~{n}eda Priego for helpful comments. Financial support by PIFI-2011, PIEC, PROMEP (1035/08/3291), and CONACyT (through the Red Tem\'atica de la Materia Condensada Blanda) is kindly acknowledged.
\end{acknowledgements}

\appendix

\setcounter{section}{0}

\section{Some important theorems and identities}
\subsection{Boundary terms and Green formula}
\noindent The boundary terms in proposition (\ref{prop1}) can be re-written as follows
\begin{eqnarray}
J^{a}_{k}=\sum^{k-1}_{j=0}\left\{\frac{1}{D^{j}}\frac{\partial^{j}}{\partial t^{j}}\left(\Delta^{k-j-1}_{g}\mathcal{O}\right)\nabla^{a}P-\frac{1}{D^{k-j-1}}\frac{\partial^{k-j-1}}{\partial t^{k-j-1}}P\nabla^{a}\Delta^{j}_{g}\mathcal{O}\right\}.
\end{eqnarray}
Remark that this expression involves the terms $P$ and $\nabla^{a}P$, then by imposing the mixed Neuman and Dirichlet boundary conditions $\left.P\right|_{\partial\mathbb{M}}=0$ and $\left.\nabla^{a}P\right|_{\partial\mathbb{M}}=0$ we are able to ignore the second term of equation (\ref{result0}). 

\noindent A key ingredient for the proof of the proposition (\ref{prop1}) is the Green formula. Let $\phi_{1}$ and $\phi_{2}$ two scalar function on the manifold $\mathbb{M}$ then the following identity is satisfied
 \begin{eqnarray}
 \label{GreenFormula}
 \int_{\mathbb{M}} dv ~\phi_{1}\Delta_{g}\phi_{2}=\int_{\mathbb{M}} dv~\left(\Delta_{g}\phi_{1}\right)\phi_{2}+\int_{\partial \mathbb{M}} da \left(\phi_{1}\partial_{\nu}\phi_{2}-\left(\partial_{\nu}\phi_{1}\right)\phi_{2}\right), 
 \end{eqnarray}
 where $da$ is the volume element of the boundary, $\partial_{\nu}={\nu}\cdot{\nabla^{a}}$ and $\nu$ is the outer normal vector of the boundary $\partial{\mathbb{M}}$.

\subsection{One real variable theorems}
\noindent In this subsection let us follow \cite{Spivak1}. Let $f:I\subset\mathbb{R}\to\mathbb{R}$ a function. Let us denote $f^{\left(n\right)}\left(t\right)\equiv d^{n}f\left(t\right)/dt^{n}$ and $a_{k}=\frac{f^{\left(k\right)}\left(a\right)}{k!}$. In addition, let us define the Taylor polynomia
\begin{eqnarray}
P_{n,a}\left(t\right)=a_{0}+a_{1}\left(t-a\right)+\cdots+a_{n}\left(t-a\right)^{n}.
\end{eqnarray}

{\theorem\label{Taylor}(Taylor theorem). Let us suppose that $f^{\prime}, \cdots, f^{n+1}$ are defined and integrable in $\left[a, t\right]$, and that the remainder $R_{n, a}\left(x\right)$ is defined by 
\begin{eqnarray}
f\left(t\right)=f\left(a\right)+f^{\prime}\left(a\right)\left(t-a\right)+\cdots+\frac{1}{n!}f^{\left(n\right)}\left(t-a\right)^{n}+R_{n,a}\left(t\right),
\end{eqnarray} }
then
\begin{eqnarray}
R_{n,a}\left(t\right)=\int^{t}_{0}\frac{\left(t-\tau\right)^{n}}{\left(n+1\right)!}f^{\left(n+1\right)}\left(\tau\right).
d\tau
\end{eqnarray}

{\theorem\label{Teo1} Let $f:I\subset\mathbb{R}\to\mathbb{R}$ a function for which $f^{\prime}\left(a\right), \cdots, f^{n+1}\left(a\right)$ with $a\in I$ exist, then 
\begin{eqnarray}
\lim_{t\to a}\frac{f\left(t\right)-P_{n,a}\left(t\right)}{\left(t-a\right)^{n}}=0
\end{eqnarray}
}

\noindent In the sense of this theorem for the limit where $t\to a$ we write $f\left(t\right)\approx P_{n, a}\left(t\right)$.

\subsection{Geometry identities}\label{ap}

\vskip0.5em
\noindent The following identities are useful for the calculations of the mean-values. This identities can be straighforward find them using the Weingarten-Gauss, (\ref{WG1}) and (\ref{Weingarten-Gauss}), several times .
\begin{eqnarray}
\label{id3}
\nabla_{a}\left(K{\bf N}\right)&=&\left(\nabla_{a}K\right){\bf N}+KK_{ab}~{\bf e}^{b}\\
\label{id4}
\Delta_{g}{\bf N}&=&\nabla_{a}K^{ab}{~\bf e}_{b}-K^{ab}K_{ab}{~\bf N}\\
\label{id5}
\Delta_{g}\left(K{\bf N}\right)&=&\left(\Delta_{g}K-KK_{ab}K^{ab}\right){\bf N}+\left(K\nabla_{a}K^{ab}+2K^{ab}\nabla_{a}K\right){\bf e}_{b}\\
\label{id6}
\nabla_{c}\Delta_{g}\left(K{\bf N}\right)&=&\left\{\nabla_{c}\left(\Delta_{g}K-KK_{ab}K^{ab}\right)-\left(K\nabla_{a}K^{ab}+2K^{ab}\nabla_{a}K\right)\right\}{\bf N}\nonumber\\
&+&\left\{\left(\Delta_{g}K-KK_{ab}K^{ab}\right)K_{c}^{~d}+\nabla_{c}\left(K\nabla_{a}K^{ab}+2K^{ab}\nabla_{a}K\right)K^{ad}\right\}{\bf e}_{d}\\
\end{eqnarray}
In particular, identities (\ref{id3})-(\ref{id6}) are useful to determine the mean and mean-square Euclidean displacement $\delta{\bf R}$. Now, for the projected displacement, $\delta{\bf R}_{\perp}$ are useful the following identities.
\begin{eqnarray}
\label{id1}
\nabla_{a}\delta{\bf R}_{\perp}&=&{\bf e}_{a}-\nabla_{a}h\hat{\bf k}\\
\label{id2}
\Delta_{g}\delta{\bf R}_{\perp}&=&-\left(K{\bf N}+\Delta_{g}h~\hat{\bf k}\right)
\end{eqnarray}

\end{document}